\documentstyle[aps,preprint,eqsecnum,epsf]{revtex}

\def\bea{\begin{eqnarray}}
\def\eea{\end{eqnarray}}
\def\bce{\begin{centering}}
\def\ece{\end{centering}}
\def\bit{\begin{itemize}}
\def\eit{\end{itemize}}
\def\slashchar#1{\setbox0=\hbox{$#1$}           
   \dimen0=\wd0                                 
   \setbox1=\hbox{/} \dimen1=\wd1               
   \ifdim\dimen0>\dimen1                        
      \rlap{\hbox to \dimen0{\hfil/\hfil}}      
      #1                                        
   \else                                        
      \rlap{\hbox to \dimen1{\hfil$#1$\hfil}}   
      /                                         
   \fi}                                         %
\newcommand\email[1]{\footnote{#1}}

\def\tightenlines{\def\baselinestretch{1.2}\small\normalsize}
\tightenlines

\begin{document}
\draft
\preprint{UCD-99-22}
\title{Measuring Gaugino Soft Phases and the LSP Mass At Fermilab}
\author{S. Mrenna,\email{mrenna@physics.ucdavis.edu}\\
\it Physics Department, University of California at Davis, Davis, CA  95616, USA}
\author{G.L. Kane\email{gkane@umich.edu} and Lian-Tao Wang\email{liantaow@umich.edu}\\
\it Randall Laboratory, Dept. of Physics, Ann Arbor, MI  48109, USA}

\date{\today}
\maketitle

\begin{abstract}
Once superpartners are discovered at colliders, the next challenge will be to
determine the parameters of the supersymmetric Lagrangian.
We illustrate how the relative phases of
the gluino, $SU(2)$, and $U(1)$ gauginos and the Higgsino mass parameter $\mu$
can be measured at a hadron collider
without ad hoc assumptions about
the underlying physics,
focusing on Fermilab.  We also
discuss how the gluino and LSP masses can be measured.
\end{abstract}

\newpage

\section {Introduction}
        Presently there is considerable indirect evidence that the
Standard Model (SM) is extended to a supersymmetric SM (SSM), and that at
least some superpartners are light enough to be produced at the
LEP or Fermilab colliders.
There is some motivation that gluinos are light and will be
produced copiously at Fermilab in the next run \cite{Kane:1998im},
based on the success of supersymmetry in explaining electroweak symmetry
breaking, which relates superpartner masses to the $Z$ boson mass.
Once superpartners are discovered, the
next step will be to measure the parameters of the Lagrangian, and then to
learn the underlying theory that leads to such a Lagrangian (technically,
the softly broken supersymmetric Lagrangian).

Before using existing information to constrain the Lagrangian, 
the minimal case has over 100 free parameters \cite{Dimopoulos:1995ju}, 
of which at most
two are directly measurable, the gluino and gravitino masses, since these
are the only ones that do not mix to form mass eigenstates.
The parameters consist of (at least) 33 masses, 40 phases, and 32 super-CKM
type angles relating flavor and mass eigenstates.  The phases can all
induce CP violating effects, and it has recently been argued that all CP
violation could be explained by these SUSY soft phases \cite{Brhlik:1999hs}.  
The phases affect
much more than CP violating observables -- cross sections, branching fractions,
radiative corrections, the LSP relic density, LEP Higgs boson limits, etc.
\cite{Brhlik:1998gu,fourp,Choi:1999cc}.  
In general, superpartner production cross sections, decay rates and kinematic
distributions depend on the phases, as we illustrate in this paper.  The same
gluino phase $\phi_3$ that we consider is also measurable in CP violation in
the Kaon system \cite{Brhlik:1999hs}.

        In the past \cite{Dugan:1985qf}, 
it was argued that the supersymmetric soft phases
must be small because they would otherwise induce neutron and electron 
electric dipole
moments larger than the experimental upper limits .  However, recently it
has been shown \cite{Ibrahim:1998je,Brhlik:1998zn,Pokorski:1999hz}
that large phases are not disallowed by current data provided
that reasonable relations exist among them.  In
fact, one expects relations among parameters in typical theories.
For example, a model
produced by embedding the MSSM on a particular configuration of D-branes
naturally has large phases and small EDMs \cite{Brhlik:1999ub}.  
The phases of the supersymmetry Lagrangian may or may not be
small -- it will be necessary to measure them to find out.

        In this paper we illustrate how the presence of
phases affect observables and can
be measured in gluino production and decay.  The gluino
production cross section does not depend on the phases,
but the gluino decay does.
To understand the essence of what happens, we consider the tree level
production of gluino pairs followed by the 
decay $\tilde{g} \rightarrow 
q\bar{q} \widetilde N_1$, where $\widetilde N_1$
is the lightest neutralino and the LSP.  The generalization to other production and decay channels is
straightforward but complicated.  For simplicity, we assume that
squarks are significantly heavier than gluinos, so that squarks decouple
from the production cross section.

        The phases of interest
enter initially in the gaugino and higgsino mass parameters $M_3 e^{i \phi_3}$,
$M_2 e^{i \phi_2}$, $M_1 e^{i \phi_1}$, and $\mu e^{i \phi_{\mu}}$ in the 
Lagrangian.  For our example,
the gluino phase enters directly, and the others through the neutralino
mass matrix

\begin{equation}
\mathbf{Y} = \left(\begin{array}{cccc}
M_{1}e^{i\phi_{1}}&0& -M_{Z}s_{\beta}s_{W}&M_{Z}c_{\beta}s_{W}\\
0&M_{2}e^{i\phi_{2}}& M_{Z}c_{\beta}c_{W}& -M_{Z}c_{\beta}c_{W}\\
-M_{Z}s_{\beta}s_{W}&M_{Z}s_{\beta}c_{W}&0&-{\mu}e^{i\phi_{\mu}}\\
M_{Z}c_{\beta}s_{W}&-M_{Z}c_{\beta}c_{W}&-{\mu}e^{i\phi_{\mu}}&0\\
\end{array} \right),
\end{equation}
where our notation is $s_\beta = \sin\beta, c_\beta=\cos\beta,
s_W=\sin\theta_W$ and $c_W=\cos\theta_W$.
This matrix can be diagonalized by a $4{\times}4$ unitary matrix
$\mathbf{N}$, i.e. $\mathbf{Y_D}=\mathbf{N^\dagger Y N}$. 
U(1) symmetries of the Lagrangian allow 
reparameterizations \cite{Dimopoulos:1995ju,Brhlik:1998zn,Dimopoulos:1996kn}, 
so that only the combinations
$\phi_3 - \phi_2$, $\phi_3 - \phi_1$, and $\phi_3+\phi_{\mu}$ are invariant and,
thus, observable.  

        For the case considered here, we will see that only one phase can be
measured, $\phi_{\rm eff}$, which depends on all of these Lagrangian phases (and also on
$\tan\beta$ and masses).  When additional decay channels are included,
different combinations of the Lagrangian phases can be measured.
Measurements at Fermilab alone could establish that $\phi_{\rm eff}$ was non-zero,
and therefore that at least one phase was non-zero.  That
would mean that at least one phase was fairly large in order to give a
significant numerical effect.  It is more difficult to show that the phases are
small, since some combinations of them may be small but not others.
A combination of measurements from electric dipole moments, $b$-factories,
and high energy colliders producing superpartners would establish that the
phases were small if they indeed were.

        It is important to emphasize that we are following a general
procedure, with no model dependent assumptions about phases being small
or related to one another, or masses being degenerate, etc.  Our simplifying
assumptions are used only to present the results simply; the procedure can
be followed without simplifying assumptions once real data once 
is available.
        
   In addition to phases, two quantities are particularly important to measure because they
affect many other results.  One is the LSP mass.  The LSP is a very good
candidate for the cold dark matter of the universe.  If superpartner
events
are observed with an escaping LSP it will, of course, greatly encourage
us to indeed believe the LSP is the CDM.  But that is not established
until a calculation of the relic density is carried out that shows
$\Omega_{LSP} \sim  1/4$, and the calculation cannot be done \cite{Brhlik:2000}
until the LSP mass and
couplings are measured and perhaps some superpartner masses.
If the LSP candidate interactions are observed in
explicit dark matter detection experiments, the LSP mass deduced from those
can be compared to the LSP collider mass.  In order to measure the LSP
mass we provide a two step procedure.

        The second important quantity is $\tan \beta$, but
it is very difficult to measure.  Here, it occurs in the $\widetilde N_1$
couplings (see next section) to quark-antiquark pairs,  which enter into 
$\phi_{\rm eff}$ along
with  phases and masses.  To the best of our knowledge,
$\tan\beta $ can only be definitively measured at
a lepton collider with a polarized beam and sufficient energy to
produce some superpartners \cite{Brhlik:1998gu}.

        In the next section we explain the formalism and derive the
relevant distributions.

\section{Relevant Lagrangian and Analytical Results}

The general gluino mass term in the MSSM Lagrangian is 

\begin{eqnarray}
{\mathcal{L}}\supset - \frac{1}{2}(M_{3} e^{i\phi_3}
\lambda_{g}\lambda_{g}+ M_3 e^{-i\phi_3}\bar\lambda_{g}\bar\lambda_{g}),
\end{eqnarray}
where $\lambda_g$ is a Majorana spinor.  With the
field redefinition $\psi_g=G\lambda_{g}$, $G\equiv e^{i\frac{\phi_3}{2}}$, the
kinematic mass of the gluino
is real.
After this redefinition, the quark-squark-gluino vertex is 
\begin{eqnarray}
{\mathcal{L}}_{q\tilde{q}\tilde{g}}= -g_s \sqrt{2} T^{a}_{jk}
(G^{\ast}\overline{\tilde{g}}_a P_L q^{k}_i \tilde{q}^{j\ast}_{iL} -
G^{\ast}\overline{\tilde{g}}_a P_R q^{k}_i \tilde{q}^{j\ast}_{iR} + 
G \overline{q}^j_i P_L \tilde{g}_a \tilde{q}^{j\ast}_{iL} -
G \overline{q}^j_i P_R \tilde{g}_a \tilde{q}^{j\ast}_{iR}),
\label{gluino_squark}
\end{eqnarray}
where the lower index $i$ is the flavor label, and the  upper indices $j$ and $k$ label the
color of (s)quarks. 

The quark-antiquark-neutralino vertex is
\begin{eqnarray}
{\mathcal{L}}_{q\overline{q}\tilde{N}_1} = -g \sqrt{2} f^{L}_i 
\overline{q}_i P_R \tilde{N}_1 \tilde{q}^{j}_{iL} + g \sqrt{2} {f^{R}_i}^{\ast}
\overline{q}_i P_L \tilde{N}_1 \tilde{q}^{j}_{iR} + h.c.,
\end{eqnarray}
where
\begin{eqnarray}
f^L_i = T_{3i}N_{12}-\tan\theta_W (T_{3i}-e_i)N_{11}~~~~~~
f^R_i = \tan\theta_W e_i N_{11}.
\end{eqnarray}  
The 
$N_{ij}$ are the complex elements of the unitary matrix which diagonalizes the neutralino mass matrix,
and depend on the phases $\phi_{1},\phi_2,\phi_\mu$.

The production of gluino pairs
does not depend
upon the phases, and can be calculated using the standard CP--conserving
Feynman rules of the MSSM \cite{Haber:1985rc}.  In the limit that the squark
masses are very heavy, the spin structure is analogous to that of top quark
pair production \cite{Mahlon:1996zn}.  The details of this calculation are
contained in the appendix.  We have included this spin structure in
our calculation to test whether it influences any physical observable.
In the end, we observed no spin correlations, so the calculation is only
mildly interesting.

Apart from an overall color factor,
the polarized decay amplitude squared for a gluino with 
spin vector $s$ is:
\begin{eqnarray}
\label{pol_decay}
|\tilde{g}(s)\to q_i\bar q_j N_1|^2 = |0|_L^2
+2\delta_L \bigg\{ m_{\tilde g} (|\alpha_L|^2 p_1\cdot p_j s \cdot p_i 
-|\beta_L|^2 p_1 \cdot p_i s \cdot p_j) \nonumber\\
        + 2 m_1 {\rm Re}(\alpha_L \beta^*_L) (p_g \cdot p_j s \cdot p_i 
- p_g \cdot p_i s \cdot p_j) \nonumber \\ 
 -2 m_1 {\rm Im}(\alpha_L \beta_L^{\ast}) \epsilon^{\mu\nu\rho\sigma} p_{g\mu}
s_{\nu} p_{i\rho} p_{j\sigma} \bigg\}+(L \rightarrow R). &
\end{eqnarray}

The spin independent piece $|0|^2=|0|_L^2+|0|_R^2$ is
\begin{eqnarray}
\frac{1}{2} \bigg(|\alpha_L|^2  (\bar t-m_1^2)(m_{\tilde g}^2-\bar t) +
|\beta_L|^2 (\bar
u-m_1^2)(m_{\tilde g}^2-\bar u) \nonumber \\ 
+2 {\rm Re}(\alpha_L \beta_L^*) m_1 m_{\tilde g} \bar s\bigg) + (L \rightarrow R),
\label{spin_ave}
\end{eqnarray}
with the couplings
\begin{eqnarray}
\alpha_L = 2 g_s g G f^{L*}_i/M_L^2~~~~~\alpha_R = 2 g_s g G^* f^{R}_i/M_R^2
\nonumber \\
\delta_L=1, \delta_R=-1~~~~~\beta_L = \alpha_L^*~~~~~\beta_R = \alpha_R^*,
\label{alpha_beta}
\end{eqnarray}
and the kinematic variables $\bar s=(p_i+p_j)^2$, $\bar t=(p_g-p_j)^2$, and $\bar u=(p_g-p_i)^2$.
The gluino and neutralino four-momenta are denoted by $p_g$ and $p_1$, respectively,
with $p_g^2=m_{\tilde g}^2$ and $p_1^2=m_1^2$.  $p_i$ and $p_j$ denote the quark and antiquark
four-momenta, and we assume $p_i^2=p_j^2=0$.
These expressions have been summed over the spins of
the final state particles.  In the limit that the phases are zero,
these expressions reproduce the standard results.
A similar formulae was derived earlier for the three--body decay
of a heavy neutralino into a lepton--antilepton pair and a lighter
neutralino \cite{Choi:1999cc}.
In writing out the expressions (\ref{alpha_beta}), we have
assumed universality of the squark masses (except for the top squarks, which
are not important for the analysis considered here)
\begin{eqnarray*}
m_{\tilde{q}_{iL}}=M_L,  m_{\tilde{q}_{iR}}=M_R,
\end{eqnarray*}
where index $i$ runs through five flavors of squarks,
and that the
squark masses are much larger than any momentum $q^2$ exchanged in the decay,
making the replacement 
\begin{eqnarray*}
(q^2-M^2)^{-1}\rightarrow -M^{-2}. 
\end{eqnarray*}
Our methodology is valid regardless of this assumption, and it is not necessary 
for analyzing the real data.  To generalize these expressions, one would simply
replace $M_{L,R}^2$ by $M_{L,R}^2-\bar t$ in $\alpha_{L,R}$ or by $M_{L,R}^2-\bar u$
in $\beta_{L,R}$.

Eq.~(\ref{spin_ave}) can be simplified, using the fact that
$|\alpha_L|=|\beta_L|$ and $|\alpha_R|=|\beta_R|$ (which is only
true with our simplifications), to
\begin{eqnarray}
\frac{1}{2} (|\alpha_L|^2+|\alpha_R|^2) \left[(\bar t-m_1^2)(m_{\tilde g}^2-\bar t) +
(\bar u-m_1^2)(m_{\tilde g}^2-\bar u) 
+4\cos \phi_{\rm eff} m_1 m_{\tilde g} \bar s\right].
\label{ex_eff}
\end{eqnarray}
The effective phase $\phi_{\rm eff}$ is defined by 
\begin{equation}
\cos\phi_{\rm eff}=\frac{|\alpha_L|^2 \cos\phi_L + |\alpha_R|^2 \cos\phi_R} 
{|\alpha_L|^2+|\alpha_R|^2}
\label{def_eff}
\end{equation}
where $\phi_L$ is $\arg(\alpha_L)  $ and $\phi_R$ is $\arg(\alpha_R) $. 
There is a summation over the index $i$, which specifies the quantum numbers
of each squark flavor.

The expressions for the production of gluino pairs and the decay outlined
above are the starting point for our phenomenological analysis.
Focusing on the decays, we note that the effect of phases
appears in Eq.~(\ref{ex_eff}) as the coefficient of the quark--antiquark
invariant mass.  Therefore, we expect to observe sensitivity to phases
in that distribution.  A similar observation was made 
for the case 
of three-body decays of neutralinos \cite{Choi:1999cc}.
The form of Eq.~(\ref{def_eff}) 
guarantees that
phase effects will not decouple if one of the
squarks is much heavier than the other. We need not  have $M_R \sim M_L $ to  
obtain a sizable effect.

Although most generally there should be four relevant phases, by 
suitable field redefinition, it can be shown that only combinations $\phi_1 - 
\phi_3$, $\phi_2-\phi_3$ and $\phi_3+\phi_{\mu}$ are relevant for this process. 
In fact, these are all R-invariant physical
observables \cite{Dimopoulos:1995ju,Brhlik:1998zn,Dimopoulos:1996kn}, 
and therefore cannot be 
further rotated away by field redefinition. 

Finally, the phases are a manifestation of CP-violation, and their presence
should be manifest in a CP-violating observable.  One example considered here
is the contraction of particle four--momenta with the 4-dimensional Levi-Cevita
tensor.

\section{Phenomenology and measuring parameters}

We concentrate on the signature of 4 jet plus missing transverse energy
$\slashchar{E}_T$ which is naturally expected for the process considered
here.
To define the signal, we specify the following cuts, motivated by the
D\O~Run I multijet and $\slashchar{E}_T$ analysis \cite{D0analysis}:
\begin{eqnarray}
\slashchar{E}_T>75~{\rm GeV}, {\mathcal{H}}_T>100~{\rm GeV}, 
|\eta^{\rm jet}|<2.5, p_T^{\rm jet}>15~{\rm GeV}, \Delta R_{ij}>0.5, \nonumber \\
0.1< \Delta\phi_{{\rm jet},\slashchar{E}_T} < \pi-0.1, 
\sqrt{ (\pi-\Delta\phi_{1,\slashchar{E}_T})^2 + \Delta\phi_{2,\slashchar{E}_T}^2}>0.5
\label{d0_cuts}
\end{eqnarray}
where $\Delta R_{ij}$ is the $\eta-\phi$ separation of jets $i$ and $j$,
${\cal H}_T$ is the scalar sum of $p_T^{\rm jet}$ excluding the leading
jet in $E_T$, and $1$ and $2$ are subscripts denoting the leading and next-to-leading
jet in $E_T$.  The $\Delta\phi$ cuts are necessary to reduce the fake $\slashchar{E}_T$
backgrounds that arise when part of a jet is lost.
For this analysis, we ignore the fact that some background events
will pass these cuts.  Their effect can only be subtracted 
statistically, and they will degrade the results presented here.
The application of cuts biases the final event sample, and the details of
the proposed measurements presented here will depend upon them.  The 
exact cuts needed to reduce backgrounds (primarily the
cut on $\slashchar{E}_T$) will not be known until the experiments begin to
collect data at high luminosity.  These cuts are meant to be representative
of what will actually be done. Finally, we smear jet energies by
a Gaussian resolution ($\sigma_E = .80\sqrt{E}$, $E$ measured in GeV)
typical of Tevatron experiments.

Some of the measurements will benefit if the flavor of the jets can
be tagged.  If all of the jets originate from light quarks, then there
is a threefold ambiguity in assigning pairs of daughter jets to a mother gluino.
We have tested various kinematic quantities, and find that choosing
the combination which minimizes the sum of the square of the jet pair
masses does fairly
well in picking the correct one.
However, there {\it is} a substantial degradation in
the measurements with this choice.
On the other hand, if two of the jets are $b$--tagged, then the
ambiguity is substantially reduced.  We estimate that this can
retain roughly a fraction of $2\times .2\times .8\times .36 = .12$
of all events, where $.36$ is an estimate of the double $b$-tag efficiency.
Charm quark tagging is also possible.
Of course, heavy flavor decays have to be corrected for the energy lost to 
neutrinos, an effect which we do not simulate here.
Finally,
jet charge can be measured
for even light quarks, using the fact that a $u$-quark will fragment to
a leading $\pi^0,\pi^+,K^+$ with (very) roughly the same probability,
while a $\bar d$-quark will fragment to a leading $\pi^+,\pi^0,K^0$.
Therefore, if four jets are observed with leading tracks of charge
$+,-,0$ and $0$, the combinations $(+,-)$ and $(0,0)$ would be
preferred.

As explained earlier, we have made the simplification that the squark masses
are universal and heavy in order to present the
results simply.  Since the mass splitting 
$\Delta M_{\tilde g\tilde N_1} \equiv m_{\tilde g}-m_{\tilde N_1}$
can be measured in a straightforward manner, we choose the
value $145$ GeV for our numerical work,
and consider gluino masses from 200 to 350 GeV in steps of 25 GeV.
The final results will not depend dramatically on the value of 
the mass splitting, provided that it is large enough so that
jets from the gluino decay are measurable.

\subsection{Coupling Measurements}

Even with the simplifications made in this analysis, the gluino
decay distribution depends on several parameters of the MSSM Lagrangian.
From Eq.~(\ref{ex_eff}), it is clear that we are sensitive to only
an overall normalization and the relative strength of the $\bar s$ term
to $\bar t$ and $\bar u$.  We can perhaps measure the overall normalization 
through the effect of the decay width on the shape of the gluino invariant
mass distribution, but the width is much smaller than the typical energy resolution.
The branching ratio for the $q\bar q\widetilde N_1$ decay
may be inferred from the number of events observed,
but one would have to measure other decay modes to make use of
this information.
Therefore, we concentrate on methods to determine $\phi_{\rm eff}$.

\begin{figure}[!ht]
\leavevmode
\bce
\epsfxsize=15cm
\epsffile{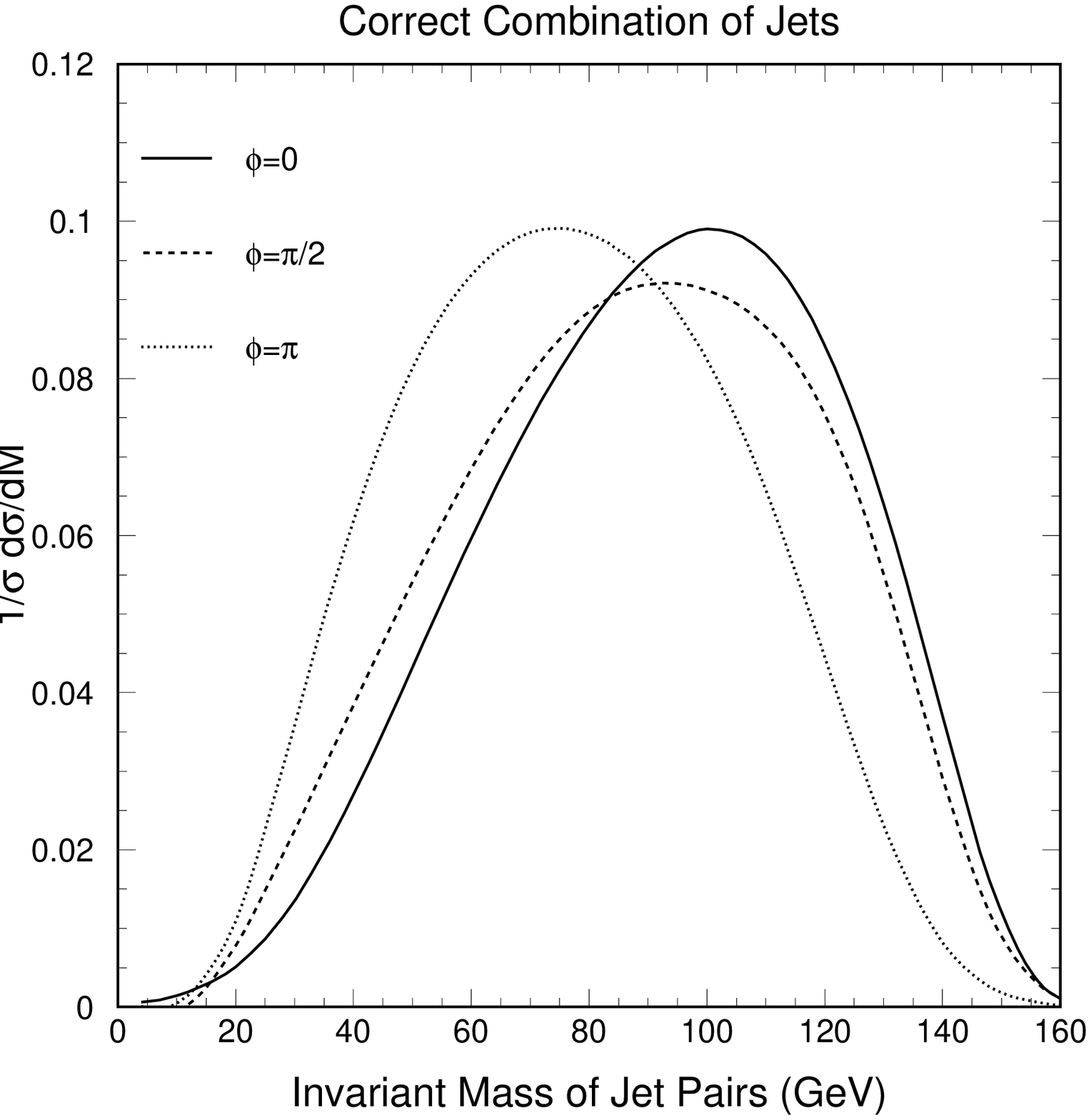}
\ece
\caption{The invariant mass distribution of jet pairs
from gluino pair production and decay at the Tevatron.
The correct combination of jet pairs is used.
The distributions for three choices of the effective
phase are shown.  A gluino mass of 250 GeV 
and a mass splitting between the gluino and LSP of 145 
GeV is assumed.  
Jet energy resolution is responsible for shifting
the endpoint upwards from 145 GeV.}
\label{mij_clean}
\end{figure}

\begin{figure}[!ht]
\leavevmode
\bce
\epsfxsize=15cm
\epsffile{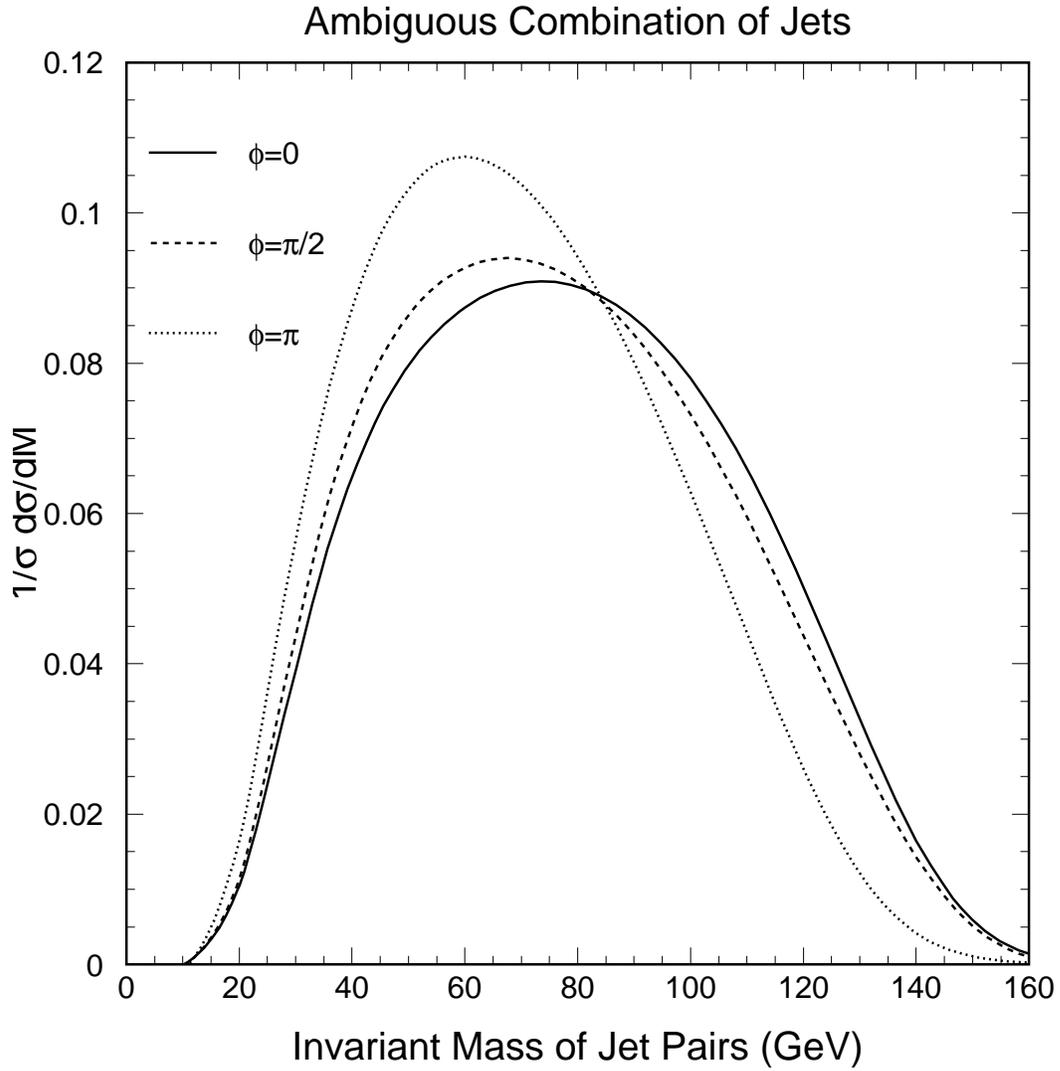}
\ece
\caption{Same as Fig.~1, except it is not assumed that
the jets will always be paired correctly.  Instead, the
combination of jet pairs that minimizes the sum of
the squared masses is used.}
\label{mij_dirty}
\end{figure}

One method is illustrated in Figs.~\ref{mij_clean} and 
\ref{mij_dirty}.  In Fig.~\ref{mij_clean}, we plot the
invariant mass distribution of the jets assuming that the jet pairs
can be unambiguously identified with a particular gluino.  
This would require, for example, that two jets have a heavy flavor
tag and two jets have an anti-tag.
The three curves (solid, long-dash, short-dash) show the effect of 
$\phi_{\rm eff}=0,\pi/2,\pi$.  From $\phi_{\rm eff}=0\to\pi$, the peak has
shifted by approximately 25 GeV.  If limited statistics require
a combination of all data, so that tagging cannot be used,
the one method for pairing the jets is to minimize the
sum of the squares of the invariant masses.  The
resultant invariant mass distribution is shown in Fig.~\ref{mij_dirty}.
There is still a phase dependence, but the miscombination has
washed the effect out somewhat (the peak has shifted roughly
10 GeV).  

To estimate how well we can actually distinguish these distributions,
we generated sets of fake data based on the distribution for
$\phi_{\rm eff}=\pi$ and compare them to the ideal distribution
for $\phi_{\rm eff}=0$.  The distribution of $\chi^2$ based
on this comparison gives
some indication of the sensitivity of the experiments.
We quote the value of $\chi^2$ that contains 5\% of the
total area of the $\chi^2$ distribution.  We could also
quote the median value of the distribution, but this would
give no indication as to the size of the low-side tail.
The interpretation of the $\chi^2$ depends upon
how the fake data was binned in a histogram.  For this case,
$\chi^2=1.9$ and 2.5 
corresponds to a 95\% and a 99\% confidence level, respectively.
Therefore, a $\chi^2$ that is far from 2.5 indicates a good
separation of the two distributions.  A $\chi^2$ near 1 means
the two distributions are consistent with each other.
We first consider the case when the two daughter jets can be paired correctly
with the mother gluino.  Assuming an efficiency of 0.2 for correctly tagging
the jet pair, $m_{\tilde g}=250 (350)$ GeV, and 2 fb$^{-1}$ of data, the comparison
yields $\chi^2=11.7 (1.3)$.  
The value $\chi^2=11.7$ for $m_{\tilde g}=250$ GeV shows that the 
two distributions are 
quite distinct with just 2 fb$^{-1}$ of data.
On the other hand,
with only 2 fb$^{-1}$ of data, the 
distribution for $\phi_{\rm eff}=\pi$ and $m_{\tilde g}=350$ GeV 
is fairly consistent with
that for $\phi_{\rm eff}=0$.  
However, with 10 fb$^{-1}$ of data, $\chi^2=6.7$.
Accepting miscombinations of the jet pairs, but an efficiency
of 1.0, the 2 fb$^{-1}$ numbers are $23.7 (2.1)$
for $m_{\tilde g}=250 (350)$ GeV.  With 10 fb$^{-1}$, the $\chi^2$ for $m_{\tilde g}=350$ GeV
increases to 11.0.
In all cases, we assume a
branching ratio of 1.0 for the four jet decay mode.
In general, we find that the increase in sensitivity scales linearly
with integrated luminosity.
Despite the fact that miscombination of the jet pairs distorts the
invariant mass distribution, a large $\phi_{\rm eff}$ dependence still
remains, so that the additional statistics makes for a better separation
of the two hypotheses ($\phi_{\rm eff}=0$ or $\pi$).

From pure kinematic considerations, 
the invariant mass distribution of the jet pairs
has an end point at $\Delta M_{\tilde g\tilde N_1}\equiv m_{\tilde g} - m_1$ 
(the end point is smeared if
we choose the wrong pairing of jets and by energy resolution).
For the measurements discussed above, and others discussed below, it is important to 
measure the endpoint with sufficient
accuracy.  
This requires a detailed knowledge of both the number of
background events and the shape of the background distribution (as
well as the same quantities for the signal).
The cuts of Eq.~(\ref{d0_cuts}) are somewhat more restrictive than
those used in the D\O Run I analysis \cite{D0analysis}, but we can
use them to estimate the background at 0.5 pb.  This means that the signal
to background ratio for $m_{\tilde g}=250$ GeV is roughly $1$,
decreasing to $1/10$ for $m_{\tilde g}=350$ GeV.  Tighter cuts
may be desired to establish a signal for heavier gluino masses and
to measure parameters.
For two very different
choices of the background shape, and for  $\cos\phi_{\rm eff}\simeq 1$ and
$m_{\tilde g}=250$ GeV,
the endpoint of the invariant mass distribution can 
be measured to a few GeV.  
There is a systematic shift in the fit endpoint and roughly twice
error for when $\cos\phi_{\rm eff}\simeq -1$, but this
is an artifact of approximating the mass spectrum by a
straight line near the endpoint.  This will not occur in
a real analysis that includes the full shape of the signal
distribution.  
While the full invariant mass distribution, including miscombinations,
gave a better separation of different parameters,
a tagged sample of events will probably yield a 
cleaner measurement of the endpoint of the invariant mass distribution.

\begin{figure}[!ht]
\leavevmode
\bce
\epsfxsize=15cm
\epsffile{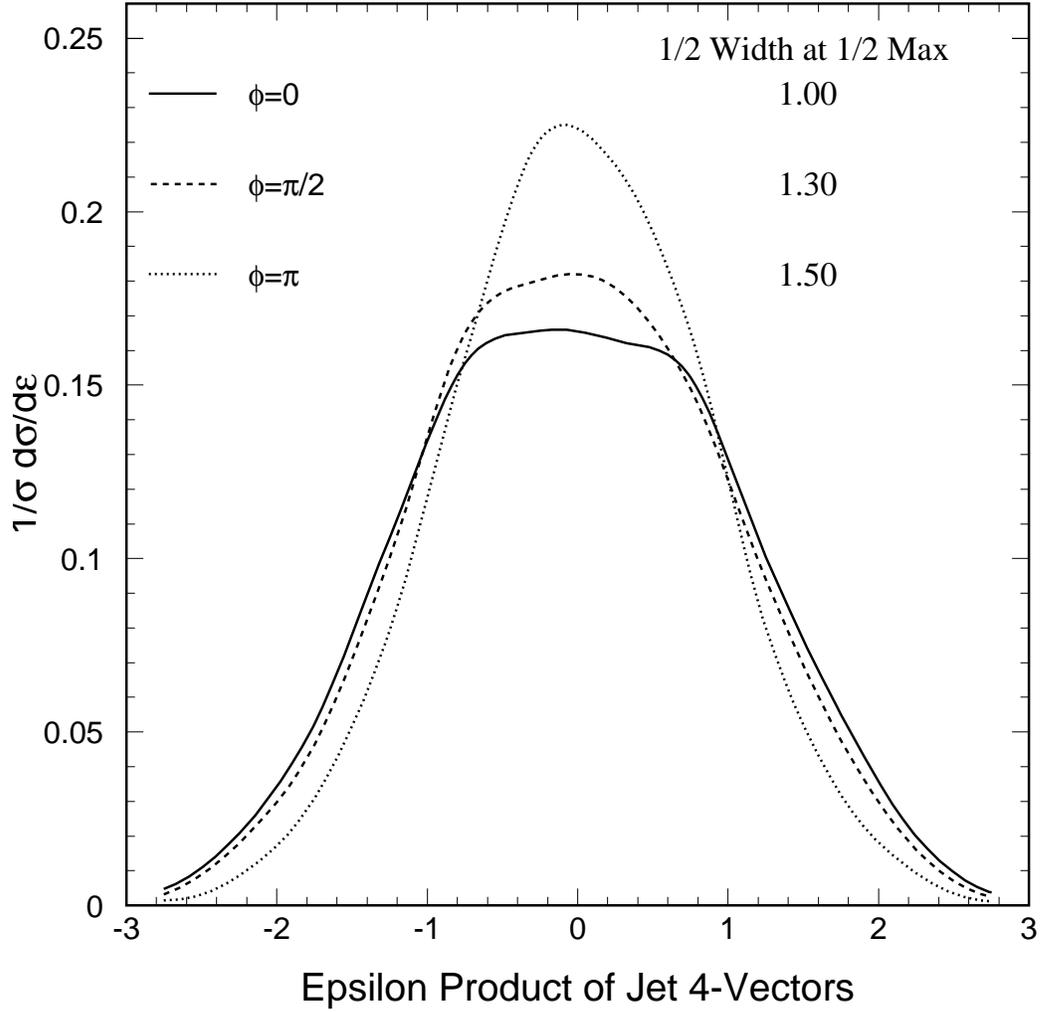}
\ece
\caption{A laboratory observable based on
contracting the four momenta of the four jets
with the Levi-Cevita tensor $\epsilon_{\mu\mu\sigma\rho}$.
The distribution is normalized by the jet energies and
is thus dimensionless.  The different
distributions and the half width at half maximum
values for three different
effective phases is shown.}
\label{eprod}
\end{figure}

One distribution that does not depend on the jet pairing or the
measurement of the endpoint
is based on contracting the jet 4-momenta with the Levi-Cevita
tensor.   We define:
\begin{eqnarray}
\epsilon = {\epsilon_{\mu\nu\rho\sigma}p_1^\mu p_2^\nu p_3^\rho p_4^\sigma
\over E_1 E_2 E_3 E_4},
\end{eqnarray}
where the momentum components are measured in the laboratory frame.  
This distribution is shown in Fig.~\ref{eprod}.  The distinguishing feature
is the half width at half maximum
for the curves, which varies from 1.0 to 1.5
as $\phi_{\rm eff}=0\to\pi$ (this is for the case of $m_{\tilde g}=250$ GeV, but
the variation for other masses is small).  We estimate the sensitivity as for
the invariant mass distribution.
For $m_{\tilde g}=250$ GeV, $\chi^2=9.3$ with only 2 fb$^{-1}$, while
for $m_{\tilde g}=350$ GeV, $\chi^2=1.0 (4.8)$ for 2 (10) fb$^{-1}$.
Therefore, the untagged jet-jet invariant mass distribution yields
a superior separation of $\phi_{\rm eff}=0$ or $\pi$ compared to any other
method considered here.  The best result will come from combining several
observables.

For Figs.~\ref{mij_clean}-\ref{eprod}, 
we have used the particular example of
$m_{\tilde g}=250$ GeV.  
The invariant mass and $\epsilon$ distributions discussed above are not
very sensitive to this number, so the
existence of CP-violating phases can be established without knowing
the masses in the problem (the mass splitting is inferred from the
endpoint of invariant mass distribution).  
In the next subsection, however, we
will demonstrate that the masses can also be known.

\subsection{Mass Measurements}

From the observed mass splitting $\Delta M_{\tilde g\tilde N_1}$, 
a measurement of $m_{\tilde g}$ will accordingly give us a 
measurement of $m_1$. 
Even in the case when the mother gluino is not tagged, comparison of different
$m_{ij}$ distributions can yield a measurement of $\Delta M_{\tilde g\tilde N_1}$.  
This is demonstrated in Figs.~\ref{mij_clean} and \ref{mij_dirty} previously mentioned.

By comparing the total number of events to the theoretical prediction
for the cross section, one can estimate $m_{\tilde g}$, and therefore determine all the
masses in the problem.  However, the total
number of events really only measures $\sigma\times BR^2(\tilde g\to q\bar q\widetilde N_1)$,
and some decay modes may not be measurable at a hadron collider.
What is preferable is
a statistical measure of the gluino mass based on kinematics.  
As a motivation, we direct the reader to the measurements of the top quark
mass in dilepton events, which can be performed even though there are two
neutrinos per event \cite{top_dilepton}.  
Given all the kinematic constraints for $t\bar t$ production and decay, there is actually
only one unknown, and the momenta of the neutrinos can be reconstructed
(there are four solutions) for an assumed top quark mass.  There is a wide
range of masses $m_t$ that yield ``reasonable'' solutions, but a probability
can be assigned to each $m_t$ by comparing the resultant kinematic distributions
to those expected for the assumed $m_t$.

The case of gluino decay is more complicated, since there are no $M_W$ constraints.
Even after specifying $m_{\tilde g}$, the longitudinal momentum $p_z$ of the LSP's is
not fixed.  We proceed by making simple fits (using our parton level
generator) to $p_z$ as a function of $m_{\tilde g}$
(the sum of a wide and a narrow Gaussian), properly normalized to unity.
The individual $p_z$ values are not noticeably correlated, and we assume they
are independent in making the fit.
For each event, a range of values for $m_{\tilde g}$ is assumed,
and an integration is performed over the two $p_z$ distributions
(the integral is replaced by a finite sum).  For each fixed value
of $p_z^1$, $p_z^2$, and $m_{\tilde g}$, the full kinematics (up to
multiple solutions) are reconstructed.  Finally, a 
Gaussian weight or probability
is assigned to each $m_{\tilde g}$ based on the difference 
$(\sqrt{\hat{s}}-2 m_{\tilde g})$/(20 GeV).

\begin{figure}[!ht]
\leavevmode
\bce
\epsfxsize=15cm
\epsffile{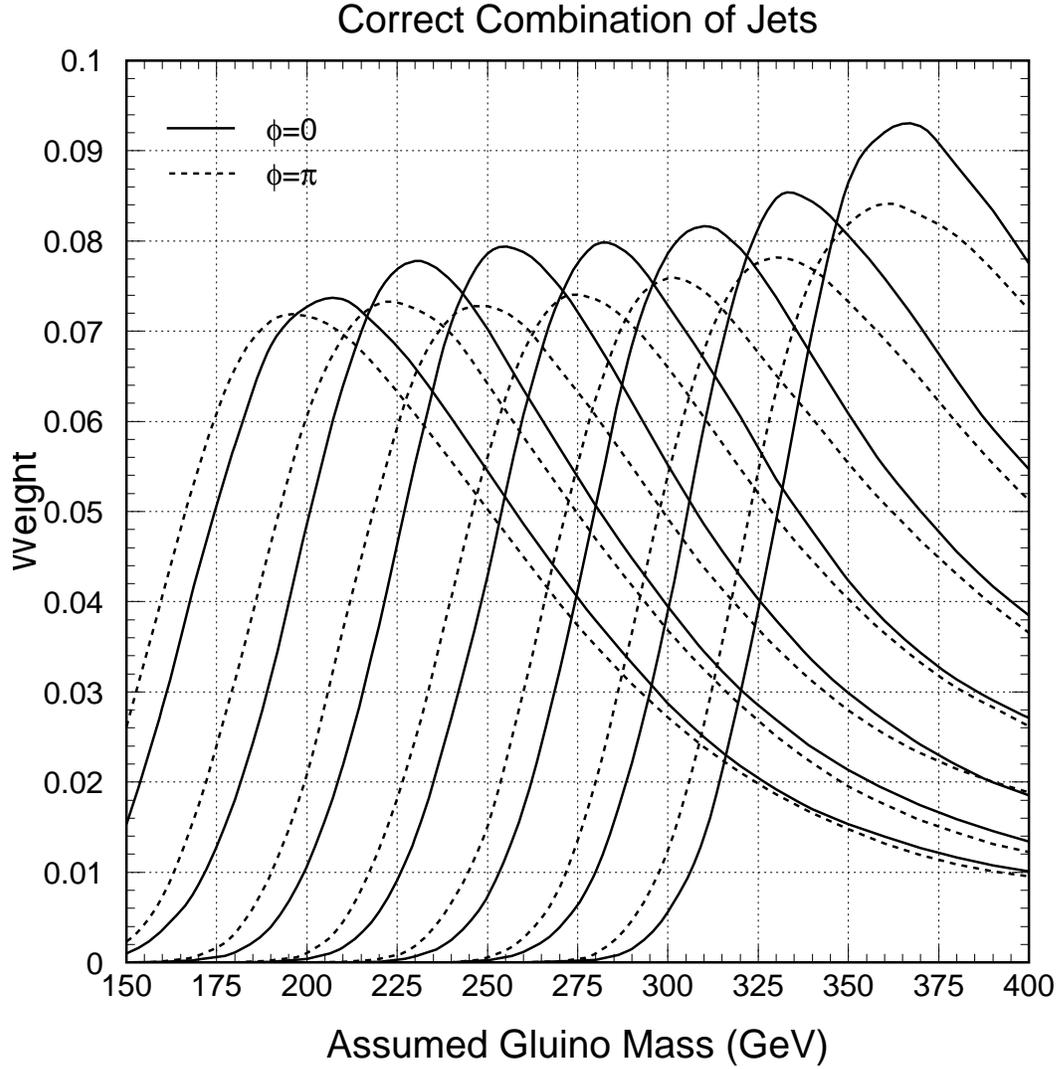}
\ece
\caption{Distribution of the probabilistic
fits to the gluino mass for $m_{\tilde g}=250-350$ GeV
in 25 GeV steps.  The peak of each distribution 
corresponds to the value of $m_{\tilde g}$ that
best describes the kinematics of four jet
plus $\slashchar{E}_T$ events.  Also shown
is the systematic shift in the distributions 
for different values of the effective phase.
The correct combination of jet pairs is used
to determine the gluino kinematics.}
\label{massfit_clean}
\end{figure}

\begin{figure}[!ht]
\leavevmode
\bce
\epsfxsize=15cm
\epsffile{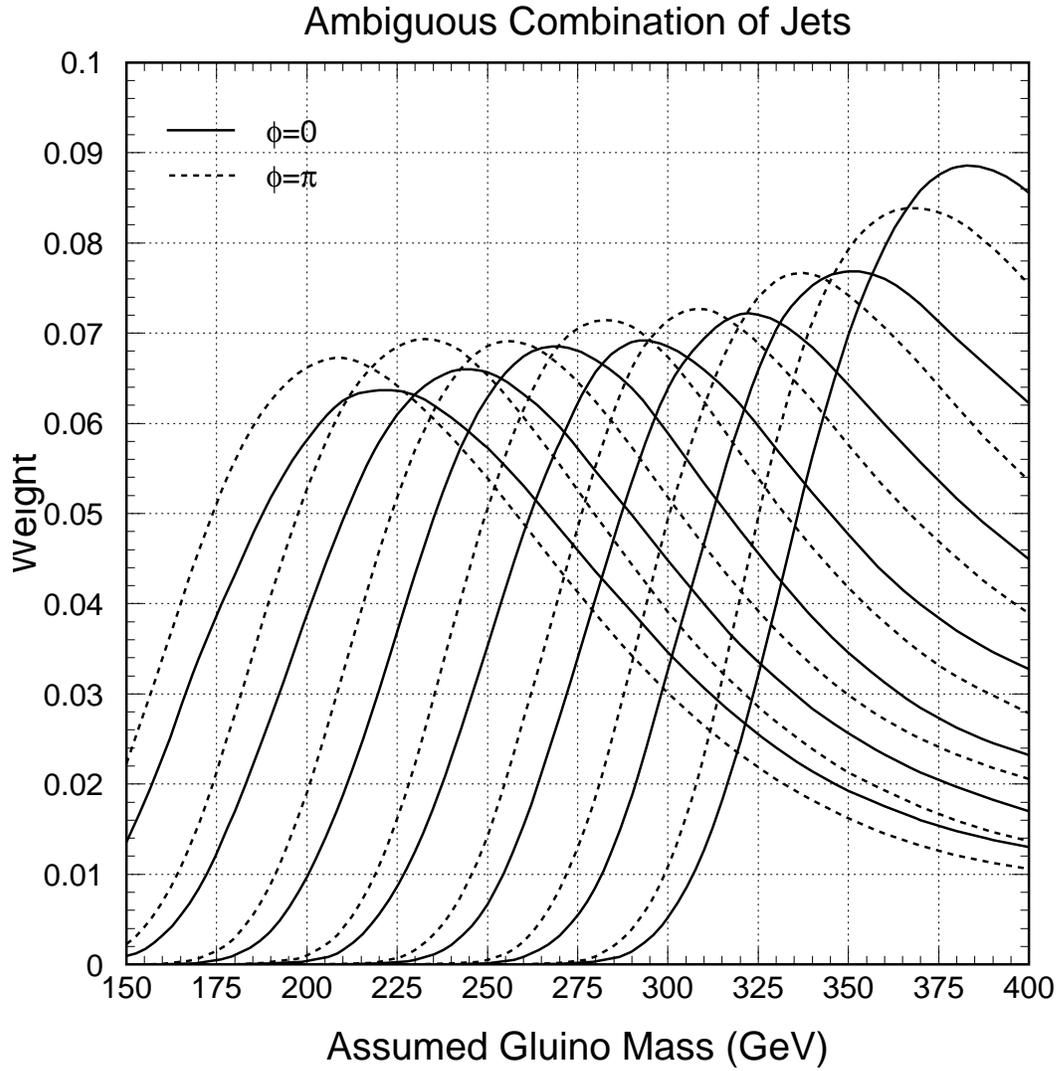}
\ece
\caption{Same as Fig.~5 but allowing for 
miscombination of the jet pairs.  When the
jets are mismatched, the gluino energy
and momentum is shifted from its actual value.}
\label{massfit_dirty}
\end{figure}

The resultant distributions as a function of $m_{\tilde g}$ are
displayed in Fig.~\ref{massfit_clean} for the correct assignment of jets
to a mother gluino (this assumes picking those events which 
satisfy a tagging as outlined above).  The same distributions
using the minimum $m^2$ algorithm are shown in Fig.~\ref{massfit_dirty}.
First, we note that we always reconstruct the correct mass
(the peak of the weight function)
within about 10 GeV
for the correct assignment of jets.  The case of $\phi_{\rm eff}=\pi$
seems to yield a larger mismeasurement, but this is not quite
correct.  For simplicity, we have used the expected $p_z$ distribution
for the case of $\phi_{\rm eff}=0$ in all cases.  However, we showed
above that the existence of a phase can be inferred independently of
the gluino mass, so the correct distribution can and should be used
for each phase.
Secondly, the reconstructed
value considering combinatoric ambiguities (Fig.~\ref{massfit_dirty}) is always roughly
25 GeV higher than the real mass, which can be corrected.  Also,
the distributions are much wider, so there will be a greater uncertainty
in the measured mass.
To estimate how well the masses can be measured, we fit a Gaussian
to the distribution for $m_{\tilde g}=250$ GeV around the peak and recorded the error on the
fit mean.  Allowing for miscombinations of the jet pair, the fit mean
was $272\pm 5 (2)$ GeV for 2 (10) fb$^{-1}$ of data.  Assuming the correct
jet pairing, but an efficiency of 0.2, the fit mean was
$259\pm 9 (4)$ GeV for 2 (10) fb$^{-1}$ of data.
 
The method outlined here is merely a simple demonstration of
principle.  While details such as soft gluon radiation,
non--Gaussian errors in energy measurements, and background
contamination would degrade our results, it is also possible
that more sophisticated pattern recognition techniques would
yield a net improvement.

\section{conclusion}

In this paper, we have considered some of the effects of CP-violating phases 
from the supersymmetry Lagrangian on
the phenomenology of gluino pair production and decay.  Such phases also
affect non--CP--violating observables.  Our specific results
are for the Fermilab collider, but the general approach applies equally well
to the future LHC collider.
We have also considered the issue of measuring sparticle masses, which is
not particular to the existence of the phases.
Although our analysis uses certain 
simplifications in order to demonstrate analytically how the
phases enter, the method we present here shows that those measurements 
can be done if we have experimental data for that process. Our method can be  
used as a prototype for more complete analyses to treat real experimental 
data.  

We concentrated on the case of gluino pair production, followed by
the gluino decay to a quark--antiquark pair and a neutralino LSP.
We found that there are simple measurements of the gluino decay products
which are sensitive to a combination of phases and couplings which
can be parametrized by an angle $\phi_{\rm eff}$.  If $\phi_{\rm eff}$
is non-zero, then it is a sign of CP violation originating in the
supersymmetry soft--breaking Lagrangian.

We have also demonstrated a method to measure the LSP mass.  This
is based on first measuring the gluino mass $m_{\tilde g}$, even though there
are two invisible LSP's in each event. The method is similar to that
used to measure the top quark mass in dilepton events: by comparing
the observed kinematics of an event to the theoretical prediction
for a range of assumed gluino masses, one can determine which
gluino mass is most consistent with the data.  In our analysis,
the most consistent gluino mass is very close to the real gluino
mass used to generate events.  Then, once $m_{\tilde g}$ is known, the
LSP mass is determined from the endpoint of the jet-jet invariant
mass.  The measured gluino and LSP mass are not the exact quantities
that appear in the MSSM Lagrangian.
The physical gluino mass $m_{\tilde g}$ can be related to the gluino mass
parameter $M_3$ through QCD, but this does require some knowledge
of the squark masses.  More measurements will be needed to
relate $m_1$ to the parameters $M_1$, $M_2$, $\mu$, and $\tan\beta$.

If Supersymmetry is responsible for stabilizing the hierarchy between
the weak and Planck scales, then there is a strong possibility that
sparticles will be produced in the near future at the Tevatron.
If the gluino is light enough, it will be produced in quantity, allowing
its discovery and, as argued in this paper, a measurement of its and other
sparticle properties.  Based on our present knowledge, these measurements
might indicate that Supersymmetry is responsible for all of CP-violation
and/or that a neutralino LSP constitutes the dark matter of the universe.

\section*{Acknowledgements}
We thank M. Brhlik, L. Everett, G. Grim, R. Breedon, and J.D. Wells
for useful comments and criticisms.

\section*{Appendix:  Calculation of Production Rates}

To maximize the sensitivity to couplings, we calculated the tree level
gluino pair production processes including the spin correlations.
We followed closely the calculations of $t\bar t$ production
in the ``off-diagonal basis'' 
\cite{Parke:1996pr,Mahlon:1996zn,Mahlon:1997uc}.
When the squark masses are heavy, one finds the same topology of
Feynman diagrams for the $q\bar q\to \tilde g\tilde g$ subprocess
as for $t\bar t$ production, and the latter results can be
used directly
with a simple modification of
the color factor.  The relevant expressions can be found in the
above references.  To summarize, in the off-diagonal
basis, when the gluinos are produced in the narrow width approximation,
only the spin combinations ($\uparrow,\downarrow$) or ($\downarrow,\uparrow$) 
occur, where the gluino spin is quantized along the special axis
that defines the basis (in the rest frame of the gluino, the spin
vector makes an angle $\pi-\xi$ with respect to the gluino direction 
of motion in the gluino pair rest frame;  The angle $\xi$ is related
to the scattering angle $\theta^*$ 
in the gluino pair rest frame and the gluino
velocity via the relation $\tan\xi=\sqrt{1-\beta^2}\tan\theta^*$).
Once the gluino decay is included, there can be interference between
the ($\uparrow,\downarrow$) and ($\downarrow,\uparrow$) amplitudes,
but these are suppressed by a factor of $\beta^2$.
To simplify our phenomenological analysis, we ignored this
interference effect.  This approximation was justified in the end,
since we observed no spin correlations between the two gluino decays.
Each gluino was decayed with a distribution given by the formulae of Sec. II.

The subprocess $gg\to \tilde g\tilde g$
has a more complicated spin structure, and is numerically less important
than the $q\bar q$ subprocess at the Tevatron for the masses considered here.
For completeness, we included this subprocess in our final results.  
The structure is almost identical to that for $gg\to t\bar t$,
except for the overall color factor and the replacement
$(9+7\cos^2{\theta^*})\to (3+\cos^2\theta^*)$ in the formulae
of Ref.~\cite{Mahlon:1997uc}.
We included the different contributions from $(\uparrow,\uparrow)$,
$(\downarrow,\downarrow)$, ($\uparrow,\downarrow$) and ($\downarrow,\uparrow$)
and added the decays ignoring the possible interference.  
Again, this is justified by the fact that none of our observables depended
significantly on the spin correlations.

For the full calculation, we used {\sf CTEQ3M} structure functions,
NLO running for $\alpha_s$, and evaluated structure functions
and $\alpha_s$ at the common scale $Q=m_{\tilde g}$.  This choice of scale
reproduces well the total cross section from a complete NLO calculation \cite{spira}.

While we observed no spin correlations in our final results, one may wonder
how higher order QCD corrections may affect them.  As observed in Ref.~\cite{mrenna},
to a very good approximation, similar NLO cross sections can be reproduced
using the tree level cross section and kinematics folded with parton
showering using a modified parton distribution function.  Since the
parton showering only boosts the gluino pair, it does not upset the
spin correlations.


\end{document}